# Effect of twin boundaries on the strength of body-centered cubic tungsten nanowires

Junfeng Cui[1], Liang Ma[2], Guoxin Chen[1], Nan Jiang[1], Peiling Ke[1], Yingying Yang[3]*, Shiliang Wang[2]*, Kazuhito Nishimura[4], Javier Llorca[5,6]*

[1] Key Laboratory of Marine Materials and Related Technologies, Ningbo Institute of Materials Technology and Engineering, Chinese Academy of Sciences, Ningbo, 315201, China
[2] School of Physics and Electronics, Hunan Key Laboratory of Nanophotonics and Devices, Central South University, Changsha, 410083, China
[3] School of Physics and Optoelectronic Engineering, Shandong University of Technology, Zibo 255000, China
[4] Advanced Nano-Processing Engineering Lab, Mechanical Engineering, Kogakuin University, Tokyo 192-0015, Japan.
[5] IMDEA Materials Institute, C/Eric Kandel 2, 28906 Getafe, Madrid, Spain
[6] Department of Materials Science, Polytechnic University of Madrid, E.T.S. de Ingenieros de Caminos, 28040 Madrid, Spain
* Corresponding author. Email: yangyingying@sdut.edu.cn (Y. Y. Y.); shiliang@mail.csu.edu.cn (S. L. W.); javier.llorca@upm.es (J. L.)

**Abstract:** Twin boundaries (TBs) are assumed to be obstacles to dislocation motion and increase the strength of metals. Here, we report the abnormal phenomenon that TBs reduce the strength of body-centered cubic (BCC) tungsten (W). [1-11]-oriented W nanowires with (121) twin planes and free of dislocations were fabricated by chemical vapor deposition. *In situ* tensile tests within the transmission electron microscope were performed on single-crystal and twinned W nanowires. The fracture strength of the twinned W nanowire was 13.7 GPa, 16% lower than that of the single-crystal W nanowire (16.3 GPa). The weakening mechanism of the TB was revealed by a combination of atomic-resolution characterizations and atomistic simulations. Twinned W nanowires failed by the early nucleation of a crack at the intersection of the TB with the surface. The standard strengthening mechanism by dislocation/TB interaction was not operative in W because the high Peierls barrier and stacking fault energy in W hinder dislocation nucleation and glide. These findings provide a new insight into the influence of TBs on the mechanical properties of BCC metals.

## 1. Introduction

   Grain boundaries (GBs) are interfaces between grains with different orientations, and their effect on the mechanical properties of polycrystalline metals has been investigated for over half a century[1]. GBs can improve the strength of metals by obstructing dislocation motion[2, 3], which is responsible for the classical Hall-Petch relationship. The increase in strength with smaller grain size is often accompanied by a decrease of ductility in polycrystalline metals because dislocation storage within the grains is more limited, leading to reduced strain hardening capability and to the early development of plastic instability and the localization of deformation[4]. The Hall-Petch relationship fails when the grain size is reduced below a certain value (approximately 30 nm)[5] and GB-mediated processes dominate plastic deformation[6, 7]. Within this regime, nanotwinned Cu with very high strength and excellent ductility was reported[4, 8]. Twin boundaries (TBs) are special GBs in which the lattice structures across the TB present mirror symmetry which not only act as obstacles to dislocations but also allow to retain the strain hardening capability, leading to a desirable combination of high strength and ductility[4]. TBs have been introduced into bulk metals[9, 10], films[11], and nanowires[12] to enhance their mechanical properties. Nevertheless, most investigations of TBs are focused on face-centered cubic (FCC) metals, in which dislocations can easily move due to their relatively low stacking fault energies and Peierls barriers[13, 14]. A natural question is whether the strengthening effect of TBs is also applicable to body-centered cubic (BCC) metals which present high Peierls barriers and stacking fault energies. This answer to this fundamental question to understand the effect of TBs on mechanical properties of metals is not clear mostly due to the experimental challenge to introduce TBs in BCC metals with high stacking fault energies.

Tungsten (W) is a refractory BCC metal, which exhibits excellent mechanical properties and outstanding thermal stability, and is mainly used in high temperature structural applications in aerospace, rocket propulsion and nuclear engineering[15, 16]. TBs are uncommon in W because of its high stacking fault energy but it has been recently reported by *in situ* transmission electron microscopy (TEM) observations that twins can develop during deformation in W nanowires[17-19]. Unfortunately, the mechanical properties of the twinned W nanowires have not been investigated, partly because the deformation twins in W are unstable and detwinning occurred spontaneously after unloading, and partly because of the experimental challenges associated with the measurement of the mechanical properties of materials at the nanoscale[13, 17].

Atomic-resolution characterization of nanomaterials during deformation can be achieved by means of *in situ* mechanical tests within a transmission electron microscope (TEM)[20, 21]. Nevertheless, the cross-section of the samples is uncertain in standard *in situ* TEM tests and the stress could not be determined[22]. Micro-nano milling using conventional focused ion beam (FIB) techniques is the most popular method to prepare samples with a given cross-section that can be used to obtain quantitative information of the mechanical properties and deformation mechanisms of materials at nanoscale[23-26]. However, the mechanical properties of the material may be seriously affected by FIB-induced damage[27], and it is very difficult to achieve atomic-resolution to ascertain the deformation and damage mechanisms at the nanoscale due to FIB-induced contamination[28]. These limitations can be overcome with a recently developed method to carry out *in situ* tests of individual nanowires (NWs) in the TEM[29]. Damage and contamination induced by FIB are avoided, leading to a combination of atomic-resolution characterization and quantitative determination of the mechanical properties. Thus, W NWs in the [1-11] orientation with and without stable TBs were fabricated by chemical vapor deposition (CVD). The (121) TB planes were parallel to the axis of the NW, as confirmed by atomic-resolution TEM characterization and TEM image simulation. *In situ* tensile tests were performed in the TEM on single-crystal and twinned W NWs. It was found that the fracture strength of the twinned W NW was lower than that of the single-crystal W NW, and atomic-resolution characterizations as well as molecular dynamics (MD) simulations were used to explore the effect of TBs on the mechanical properties of the W NWs.

## 2. Materials and methods

### 2.1 Processing and characterization

W NWs were synthesized on a silicon (Si) substrate by chemical vapour deposition (CVD). $WO_3$ powders were heated at 900 ºC in an atmosphere containing water vapour and $H_2$ gas and W NWs were grown on the Si substrate. More details about the fabrication of W NWs are available in the literature[30]. W NWs were characterized by SEM (Helios G4 CX, Themo Fisher Scientific, USA), TEM (Talos F200X, Themo Fisher Scientific, USA) and aberration-corrected TEM (Spectra 300, Themo Fisher Scientific, USA). It should be noted that both twinned and single-crystal W NWs were synthesized under the same conditions and the formation mechanism of TBs in several W NWs is still unclear.

### 2.2 Sample preparation and *in situ* TEM mechanical tests

Individual W NWs were transfered and fixed with glue on push-to-pull (PTP) devices from the Si substrate in air using methodology previously developed[29]. Conductive silver epoxy glue (CircuitWorks 401 Conductive Epoxy, CW2400, Chemtronics, USA) was used to fix the NWs. Firstly, a human eyebrow hair tip was fixed on the end of a manipulator with silver paint. The Si substrate with the W NWs was put on the platform of an optical microscopy (Leica DM2500 M, Germany), and an individual W NW was picked up using the eyebrow hair tip. Then, the Si substrate was replaced by a PTP device and the NW on the eyebrow hair tip was put on it. Finally, a drop of glue was taken using the eyebrow hair tip and transferred on two ends of the NW on the PTP device under the optical microscope. Afert curing for 24 h at room temperature, the NW was strongly attached to the PTP device. *In situ* tensile tests within the TEM (Talos F200X, Themo Fisher Scientific, USA) were performed using a Hysitron PI-95 PicIndenter (PI-95 TEM PicoIndenter, Bruker, Germany) under displacement control 5 nm/s.

2.3 MD simulations

MD simulations of the tensile deformation of twinned and single-crystal W NWs were carried out using the Large-scale Atomic/Molecular Massively Parallel Simulator (LAMMPS). Periodic boundary condition were imposed on the axial direction of the NWs (X direction), while the NWs had free surfaces in Y and Z directions. The embedded atom method (EAM) potential developed by Marinica et al.[31] was employed in the simulations. This potential has been proven to well describe interactions between W atoms to simulated the deformation of single-crystal and twinned W[32, 33]. The structures of NWs were relaxed at 300 K for 100 ps using the isothermal-isobaric (NPT) ensemble and the energy of the system was minimized using the conjugate gradient scheme. NWs were stretched along X direction at 300 K at a constant strain rate of $10^9$ s$^{-1}$ in the canonical (NVT) ensemble. The time step was 1 fs in all simulations.

## 3. Results

3.1 Characterization of W NWs

   Scanning electron microscope (SEM) images of the W NWs are depicted in Fig. 1a. The inset shows that the W NW has a hexagonal cross-section. According to the crystal growth law of W NWs, the six facets of the surface belong to the {110} planes[30]. A bright field TEM image of a typical twinned W NW is shown in Fig. 1b. The (121) twin plane is perpendicular to (10-1) planes, and forms an angle of 60° with the electron-beam direction, as illustrated in the upper inset of Fig. 1b. The green-yellow schematic inserted below the NW in Fig. 1b shows that the twin plane is parallel to the axial direction of the NW and encompasses the whole NW. Detailed structural information is given by the selected area electron diffraction (SAED) pattern (Fig. 1c) and high-resolution TEM (HRTEM) images (Figs. 1d-g). The SAED pattern in Fig. 1c indicates the mirror symmetry of the matrix and the twin along {112} planes and the <111> axial orientation of the NW. The spots at the position of 1/3{112} and 2/3{112} are responsible for the double diffraction of the twin.

   High-resolution TEM (HRTEM) characterization and simulation as well as aberration-corrected high-angle annular dark-field scanning TEM (HAADF-STEM) characterization were carried out on the twinned structure to clarify the stable TB structure in the W NW. The HRTEM image (Fig. 1d) shows a band between the matrix and the twin along the NW. Structures above and below the band belong to the BCC W structure, as depicted by the enlarged HRTEM images and their corresponding fast fourier transform (FFT) patterns in Figs. 1e and 1f, respectively. The band is attributed to the overlap of the matrix and of the twin, as illustrated in the insets in Fig. 1b. The simulated HRTEM image inserted in Fig. 1g is based on a model of twinned W with a (121) twin plane viewed long the [011] orientation, and agrees well with the experimental HRTEM image taken from the band (Fig. 1g). The unoverlapped atoms of the matrix and of the twin (represented by yellow and green atoms in the model, respectively) are hard to be detected by HRTEM characterization, but they can be resolved by aberration-corrected HAADF-STEM characterization (shown in the yellow square of Fig. 1g). This image provides a strong evidence for the twin model. Furthermore, the FFT pattern (yellow square of Fig. 1d) taken from the band agrees well with the SAED pattern in Fig. 1c, also indicating the composite structure of the matrix and of the twin in the band.

### 3.2 *In situ* TEM tensile tests

In order to investigate the influence of TBs on mechanical properties of W, *in situ* TEM tensile tests were performed on single-crystal and twinned W NWs. Using a home-made setup, individual W NWs were transferred and fixed with glue on a PTP device (Fig. 2a) in air. Both single-crystal (Fig. 2b) and twinned (Fig. 2c) W NWs were stretched along the <111> direction. The inserted SAED patterns in Figs. 2b and 2c in combination with HRTEM images in Figs. 3a and 3c, evidence the single-crystal and twinned structures of the W NWs. Fig. 3b and Fig. 3d show EDS spectra taken from the single-crystal and twinned NWs, both of which indicate the pure W. The SAED pattern (inset in Fig. 2c) and the HRTEM image (Fig. 3c) of the twinned NWs are in good agreement with these of the twinned W NW showed in Figs. 1c and 1g.

The engineering stress-strain curves of the single-crystal and twinned W NWs are plotted in Fig. 2d. The curves are superposed in the elastic region, indicating that TBs have no influence on the elastic modulus of W NWs. This results is consistent with those reported in twinned Cu and InP NWs[25, 34]. The elastic modulus of 360 GPa of W NWs agrees well with that of 362 GPa reported in the literature[35]. The decrease in the slope of the black curve after the strain exceeding 2% is attributed to the thermal drift of the testing equipment, which is caused by the change of external environments. In fact, the single-crystal W NW exhibit linear elastic behavior, even the strain exceeding 3.5%, as evidenced by the stress-strain curve in Fig. 4a and snapshots of TEM images (Fig. 4b) extracting from Movie 1. The twinned W NW fractured at a strain of 4.0% with a fracture strength of 13.7 GPa, which is lower than that of the single-crystal W NW (16.3 GPa at a strain of 5.2%). Insets with black and red squares depict TEM images showing cross-sections of the single-crystal and twinned W NWs, respectively. Taking the cross-section area $S$ of the NWs into $S = \frac{\pi}{4}d^2$, we get the effective radius $d$ of 99.4 nm and 103.0 nm for the single-crystal and twinned W NWs. Thus, the size effect on mechanical properties and deformation mechanisms of the NWs can be ignored[25]. It also should be noted that the twin plane was parallel to the loading direction, and TB gliding and migration were avoided. This means that TB gliding and migration are not responsible for the TB weakening effect in W. Figs. 2e and 2f present the fracture processes of the single-crystal and twinned W NWs, extracted from Movie 2 and Movie 3, respectively. The single-crystal W NW fractured abruptly and both sections of the NW flew away (Fig. 2e), while the twinned W NW fractured into two parts but the sections did not fly away (Fig. 2f). The reason for the different fracture behavior of the single-crystal and twinned W NWs will be discussed in the following part of MD simulations.

HRTEM characterization of fracture surfaces was carried out to shed light on the fracture mechanisms of the single-crystal and twinned W NWs. An enlarged TEM image from the fracture surface of the single-crystal W NW is shown in Fig. 5a. It can be seen that plastic deformation occurred near the fracture surface. The Moiré fringes (marked by yellow arrows in Fig. 5b) indicate tilted partial lattices, as evidenced by the FFT pattern (the inset in Fig. 5b). Fig. 5c depicts a filtered image showing only (01-1) planes from the area marked by the green square in Fig. 5b, showing plenty of dislocations and slip bands. The accumulation of dislocations indicates that plastic deformation of the single-crystal W occurred before fracture and this behavior is consistent with that induced by compression in W micropillars[36] and NWs[37]. Nevertheless, dislocations, slip bands and tilted lattices are not observed in the HRTEM image (Fig. 5e) from the fracture surface of the twinned NW.

## 3.3 MD simulations

MD simulations were conducted on W NWs with and without TBs to verify the weakening effect of TBs in W observed in experiments and explore the weakening mechanism. The atomic model of the twinned W NW with a (121) twin plane is shown in Fig. 6a. The diameter of the NW is 21.8 nm, and the cross-section (the inset on the upper left in Fig. 6a) has the same shape as that of the twinned NW in the tensile test (the inset with a red square in Fig. 2d). When viewed along the [011] direction, the atomic model agrees well with the TEM image, as shown in Fig. 3c. An atomic model of the single-crystal W NW with the same cross-section and diameter (the inset with a black square in Fig. 6b) was built for comparison. When stretched along the [1-11] direction, the single-crystal W NW fractured at a strain of 13.7% with a fracture strength of 41.3 GPa, which agree well with the value of 42 GPa[33]. The fracture strength of the twinned W NW (39.8 GPa at a strain of 11.1%) was lower than that of the single-crystal W (Fig. 6b), in agreement with the experimental results presented above. It should be noted that the cross-section shapes of the single-crystal and twinned W NWs in our experiments are different (insets in Fig. 2d). Thus, another single-crystal W NW with a different cross-section shape (the inset with a green square in Fig. 6b) was built, and the MD simulation of the tensile behavior was carried out. As shown in Fig. 6b, the stress-strain curves of the single-crystal NWs with different cross-section were identical. Fracture strengths of single-crystal and twinned W NWs with different diameters are depicted in Fig. 7, and the twinned W NWs always failed at lower stress.

The fracture process – according to the MD simulations – of the twinned W NW is shown in Figs. 8a and 8b along the [10-1] and [1-11] directions, respectively. It should be noted that the atoms in the perfect BCC lattice are removed for clarity in Fig. 8b. Obviously, the crack nucleated at the surface/TB intersection at 11.25% tensile strain (Figs. 8a2 and 8b2), and propagated into the NW from the intersection (Figs. 8a2-5 and 8b2-5). Finally, the crack propagated through the whole NW and the fracture surfaces were formed (Figs. 8a6 and 8b6). In contrast, many cracks nucleated at the surface intersection of {110} planes in the single-crystal W NW at a tensile strain of 14.05% (Figs. 9a and 9b). The preferential nucleation of the crack at the surface/TB intersection is responsible for the low fracture strength of the twinned W NW, and demonstrates the weakening effect of TBs in W.

The formation and evolution of dislocations during crack nucleation and propagation in the twinned W NW viewed along the [10-1] direction is depicted in Fig. 8c. Dislocation nucleation is not associated with crack nucleation (Fig. 8c2) and they appear during crack propgation near the crack front (Fig. 8c3). Further crack propagation leads to an increase in the dislocation density, as shown in Figs. 8c4-5. Nevertheless, most of the dislocations left the NW through the fracture surfaces and only 3 dislocation segments remain within the NW after fracture (Fig. 8c6). On the contrary, many micro-cracks (cracks 2-5 in Fig. 9a) are nucleated in the single-crystal W NW after the appearance of the first crack (crack 1 in Fig. 9a) due to the high tensile stress achieved, and many more dislocations are emitted from these cracks (Fig. 9c), as compared with the twinned W NW. In fact, 247 dislocation segments remain in the single-crystal NW after fracture (Fig. 9c6). The results of the MD simulations support the experimental observations, which show many dislocations in the single-crystal W NW, but not in the twinned W NW (Fig. 5). In addition, the elastic potential energy of the NW under tensile strain will be converted into the kinetic energy as fractured[38]. Compared with the twinned W NWs, the high fracture stress of the single-crystal W NW indicates its high kinetic energy as fractured. The micro-cracks except for the main crack and the high kinetic energy in the single-crystal W NW make it more likely to fly away than the twinned W NW as fractured, which agrees well with the phenomena observed in experiments (Figs. 2e and 2f).

## 4. Discussion

TBs are traditionally introduced into metals to enhance the strength. However, it is hard to introduce TBs into BCC W due to its high stacking fault energy. As a result, the effect of TBs on mechanical properties of W is largely unknown. Deformation twinning has been verified in nanoscale W by *in situ* TEM observations, but detwinning occurs after unloading[17], and the TBs are supposed to be not parallel to the twin plane[13]. Here, [1-11]-oriented W NWs with and without stable TBs were manufactured by CVD, and *in situ* TEM tensile tests revealed that the fracture strength of twinned W was 16% lower than that of single-crystal W (Fig. 2c). MD simulations showed that the crack nucleation occurred earlier at the intersection of the TB with the surface of the W NW, while fracture started at higher strain at the intersection of {110} planes with the NW surface in the single crystals.

It should be noted that dislocation motion is very difficult in W because of the high Peierls barrier and stacking fault energy[39, 40]. Peierls stress ($\tau_P$) and normalized Peierls values ($\tau_P/G$, where $G$ is the shear modulus), as well as stacking fault energies in W and some other typical metals[40-43] are shown in Table 1. Especially, the Peierls stress and stacking fault energy in W is over 3428 and 40 times greater than that in Cu. It is well known that the strengthening effect of TBs in FCC metals is achieved by hindering dislocation motion. This mechanism, however, should not be very effective when there are few dislocations or dislocation motion is intrinsically difficult. MD simulations revealed that dislocations were nucleated after cracking in W NWs deformed in tension and, dislocation/TB interactions did not play any role from the viewpoint of strength. This is particularly true in the case of NW (as compared with bulk samples) because the initial dislocation density is negligible[44]. In addition, the sample preparation method (that does not introduce damage and contamination in the sample) minimizes the initial density of dislocations and their interaction with TBs in the twinned and single-crystal W NW. In addition, the intrinsic weakening of TBs in some other materials (Al, Ni, Al₂O₃ and SiC) has been demonstrated by numerical simulations in the absence of dislocations[45-48]. The weakening effect of TBs on the mechanical properties may also be applicable to other BCC metals with high Peierls barrier and stacking fault energy. However, it is hard to quantitatively determine the effect of TBs on their mechanical properties due to the complex competition between the intrinsic weakening of TBs and their interaction with dislocations, as well as the challenge of introducing TBs in metals with high stacking fault energy. In-depth understanding on the weakening effect of TBs on other metals requires further experimental results and atomistic simulations.

**Table 1.** Peierls stress ($\tau_P$) and normalized Peierls values ($\tau_P/G$, where $G$ is the shear modulus), as well as stacking fault energies (SFE) in typical FCC and BCC metals[40-43].

| Crystal type | Metal | $\tau_P$ (MPa) | $\tau_P/G$ ($10^{-3}$) | SFE (mJ/m²) |
|---|---|---|---|---|
| FCC | Cu | <0.28 | <0.07 | 44 |
| | Ag | <9 | <0.035 | 18 |
| | Au | <0.9 | <0.038 | 31 |
| | Al | <1.4 | <0.056 | 146 |
| BCC | W | 960 | 5.9 | 1773 |
| | Mo | 730 | 5.4 | 1443 |
| | Nb | 415 | 8.2 | 677 |
| | Ta | 340 | 5.2 | 724 |

## 5. Conclusions

In summary, [1-11]-oriented W NWs with and without stable (121) twin planes were synthetized by CVD. Individual NWs were transferred and fixed on PTP devices without contamination and damage. Twinned and single-crystal W NWs were stretched along the [1-11] direction within the TEM. The fracture strength of the NW containing TBs was lower than that of single-crystal W NW. HRTEM showed the presence of dislocations and slip bands near the fracture surface of single-crystal W, which were not found in the twinned W NW. MD simulations revealed that twinned W

NW failed by the early nucleation and growth of a dominant crack at the surface/TB intersection. Few dislocations were nucleated during crack propagation, which left the NW through fracture surfaces. Thus, dislocations were not able to interact with the TB during deformation and the typical TBs strengthening observed in FCC metals was not operative. Our results provide a new insight into the effect of TBs on mechanical properties of W, which may be important in other BCC metals and alloys with high Peierls barriers and stacking fault energies.

**Author statement**

**Junfeng Cui:** Conceptualization, Methodology, Investigation, Writing-Original Draft, Writing-Review & Editing, **Liang Ma:** Investigation, Methodology, **Guoxin Chen:** Investigation, Funding acquisition, **Nan Jiang:** Funding acquisition, Writing-Review & Editing, **Peiling Ke:** Writing-Review & Editing, **Yingying Yang:** Investigation, Writing-Review & Editing, **Shiliang Wang:** Investigation, Methodology, Writing-Review & Editing, **Kazuhito Nishimura:** Writing-Review & Editing, **Javier Llorca:** Methodology, Writing-Review & Editing,

**Data availability**

The raw/processed data required to reproduce these findings cannot be shared at this time due to technical or time limitations.

**Declaration of competing interest**

The authors declare that they have no known competing financial interests or personal relationships that could have appeared to influence the work reported in this paper.

**Acknowledgements**

The authors acknowledge the financial supports from the Youth Innovation Promotion Association CAS (2019295), Ningbo science and technology service demonstration projects (2019F1019) and the Science and Technology Major Project of Ningbo (2018B10046).

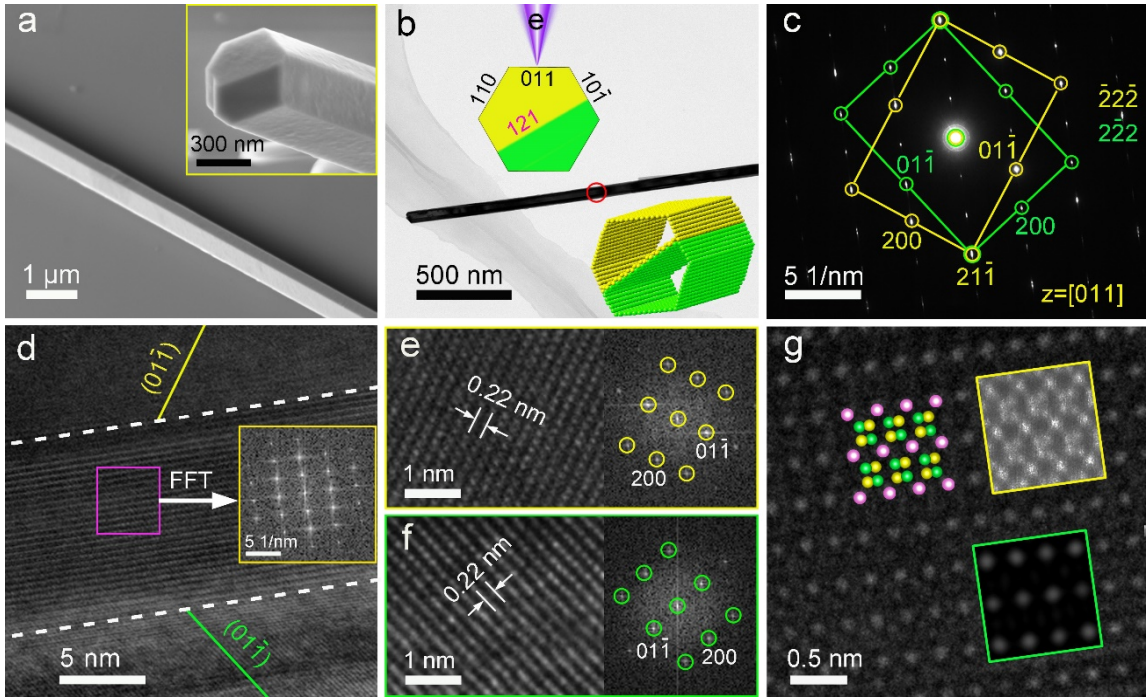

**Fig. 1.** Characterization of W NWs fabricated by CVD. (a) SEM images of W NWs. (b) Bright field TEM image of a twinned W NW. The insets above and below the NW show the cross-section and the schematic of the twinned W NW. (c) SAED pattern taken from the region marked by the red circle in (b). (d-f) HRTEM images taken from the W NW in (b). A band is located between two dotted lines in (d). The inset in the yellow square in (d) shows the FFT pattern, taken from the region marked by the pink square. HRTEM images taken from areas above and below the band are shown in (e) and (f), and the corresponding FFT patterns are shown in the insets, respectively. (g) Enlarged HRTEM image taken from the region marked by the pink square in (d). The insets in the green and yellow squares show a simulated HRTEM image based on a model of twinned W with a (121) twin plane viewed long the [011] orientation and an aberration-corrected HAADF-STEM image taken from the band, respectively. The yellow, green and pink balls of the model represent matrix atoms, twin atoms and overlapped atoms of the matrix and of the twin.

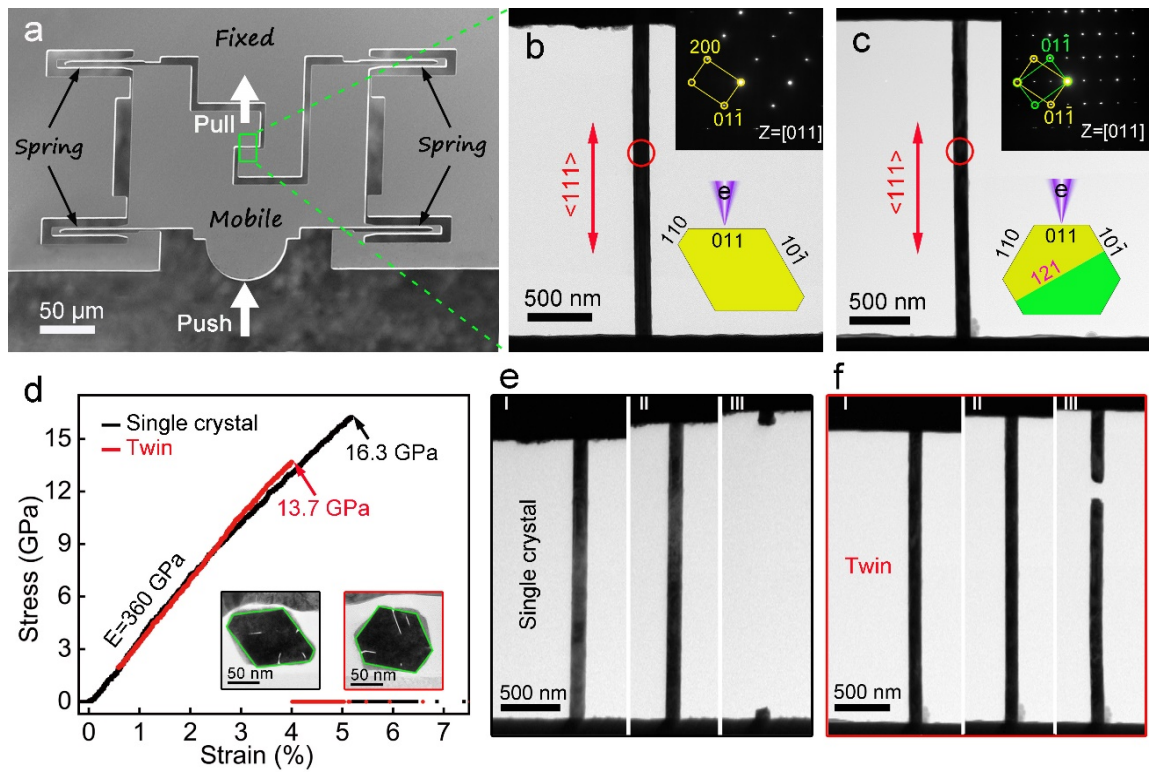

**Fig. 2.** *In situ* TEM tensile tests of W NWs without and with TBs. (a) SEM image of a PTP device used for tensile tests. (b and c) TEM images of single-crystal (b) and twinned (c) W NWs after fixed on PTP devices. The upper and lower inserts show corresponding SAED patterns taken from the regions marked by red circles and cross-section diagrams of the NWs, respectively. (d) Engineering stress-strain curves of the single-crystal (black) and the twinned (red) W NWs. Insets with black and red squares depict TEM images showing cross-sections of the NWs in (b) and (c). (**e** and **f**) Snapshots of fracture processes of the single-crystal (e) and twinned (f) W NWs.

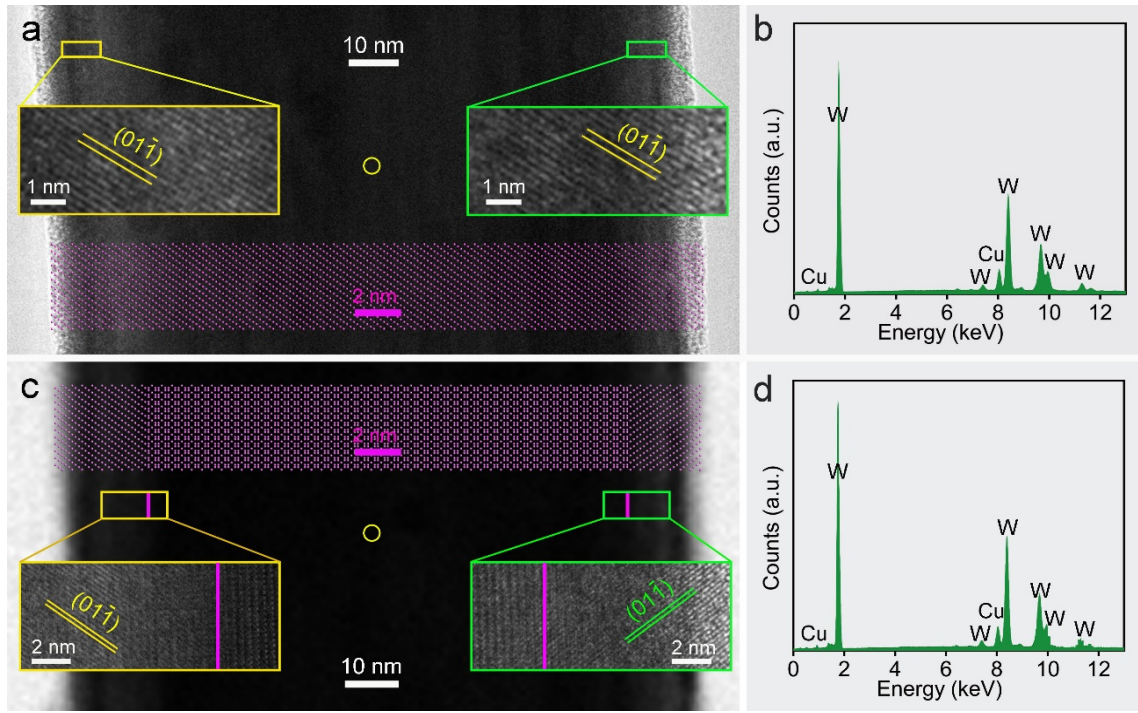

**Fig. 3.** TEM characterization of W NWs with and without TBs. TEM images of the single-crystal W NW (a-b) in Fig. 2b and of the twinned W NW (c-d) in Fig. 2c. (b) and (d) shows the EDS spectra taken from regions marked by yellow circles in (a) and (c), respectively. The insets with yellow and green squares in (a) and (c) show HRTEM images taken from regions marked by the corresponding squares. Inserted atomic diagrams show atomic models of the single-crystal W NW and of the twinned W NW with the (121) TB viewed along the [011] direction.

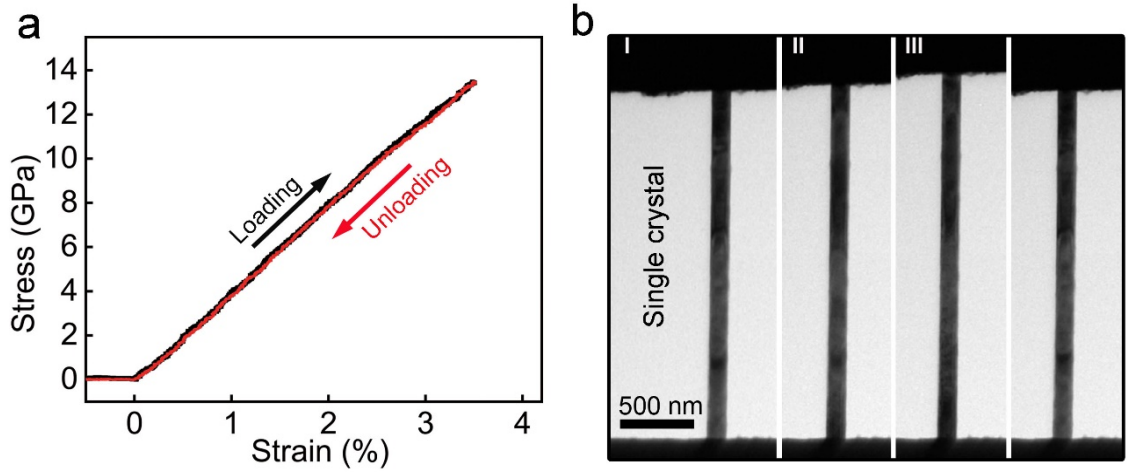

**Fig. 4.** Elastic deformation of the single-crystal W NW. The engineering stress-strain curve (a) and snapshots of TEM images (b) during loading and unloading of the single-crystal W NW in Fig. 2b.

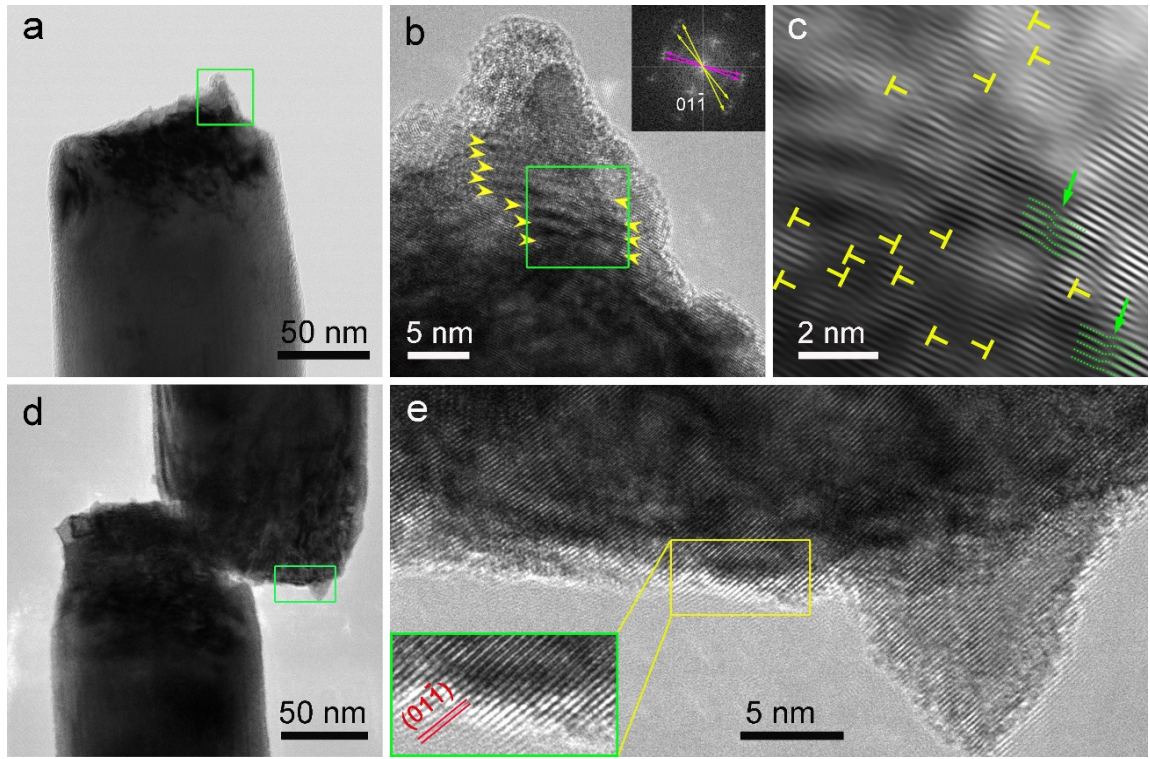

**Fig. 5.** Characterizations of the fracture surfaces of single-crystal and twinned W NWs. (a) Enlarged TEM image from the fracture surface of the single-crystal W NW in Fig. 2e. (b) HRTEM image from the region marked by a green square in (a). The inset shows the corresponding FFT pattern. (c) A filtered image showing only (01-1) planes from the area marked by the green square in (b). (d) Enlarged TEM image taken from the fracture surfaces of the twinned W NW in Fig. 2f. (e) HRTEM image from the region marked by the green square in (d). The inset shows an enlarged image from the region marked by the yellow square.

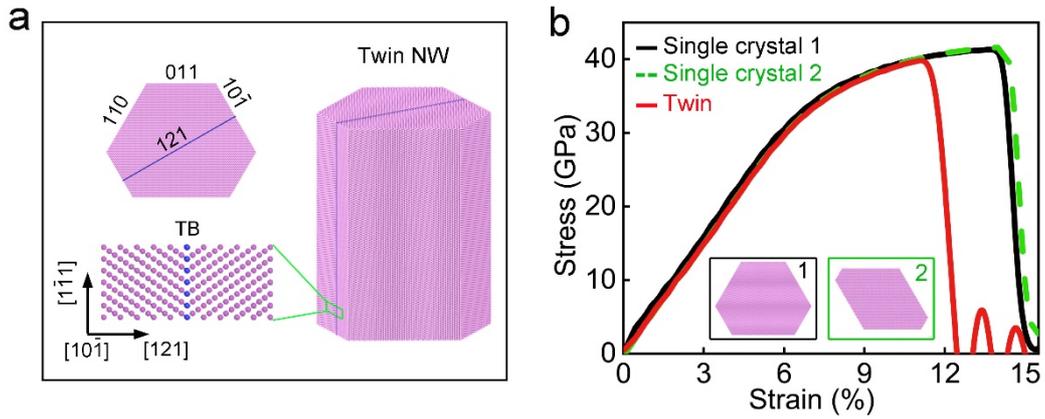

**Fig. 6.** MD simulations of the tensile tests of single-crystal and twinned W NWs. (a) The initial atomic model of a twinned W NW of 21.8 nm of diameter with a (121) twin plane along the NW. Blue atoms stand for the TB. The inset on the upper left shows the cross-section of the NW. The inset on the lower left shows the enlarged image of the atomic structure in the area marked by the green square viewed along the [10-1] direction. (b) Stress-strain curves of single-crystal and twinned W NWs deformed in tension along [1-11]. The twinned W NW is shown in (a). Insets with black and green squares show the cross sections of single crystal NW 1 and 2, respectively.

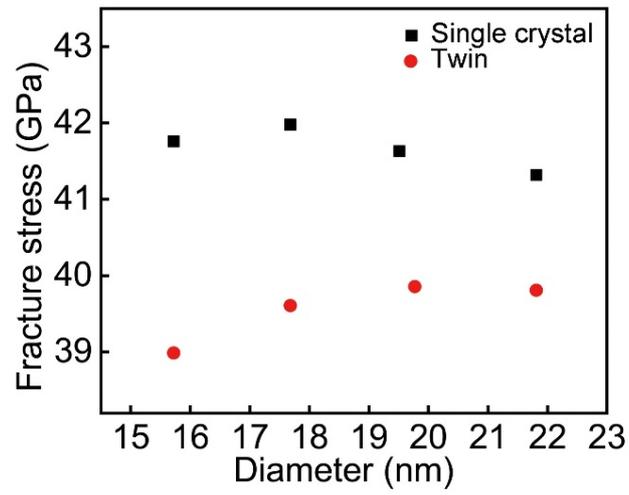

**Fig. 7.** MD predictions of the fracture strength of single-crystal and twinned W NWs deformed in tension along the [1-11] orientation as a function of the NW diameter.

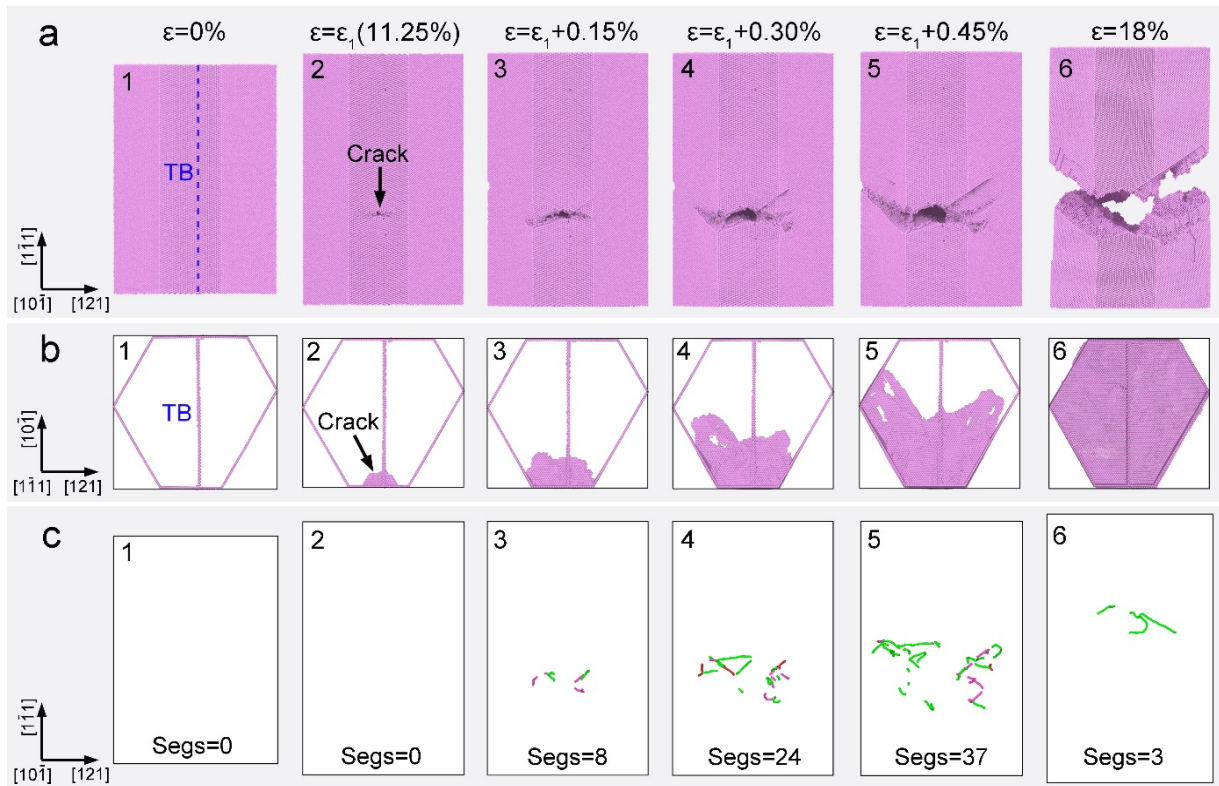

**Fig. 8.** Crack nucleation and propagation in a twinned W NW in tesnion along [1-11] direction. (a and b) Atomic model of the twin W NW with a diameter of 21.8 nm viewed along the [10-1] direction (a) and the [1-11] direction (b) at strains of 0% (1), $\varepsilon_1$= 11.25% (2), $\varepsilon_1$+0.15% (3), $\varepsilon_1$+0.30% (4), $\varepsilon_1$+0.45% (5), and 18% (6). $\varepsilon_1$ stands for the strain at the crack initiation. The blue dotted line in (a1) indicates the TB. Atoms in the perfect BCC lattice are removed for clarity in (b). (c) Dislocation configurations in the twinned NW in (a) viewed along the [10-1] direction. Green, pink and red lines represent 1/2<111>, <100> and other type of dislocations. The number of dislocation segments (Segs) is 0 (1), 0 (2), 8 (3), 24 (4), 37 (5), and 3 (6). The black frames in (b) and (c) indicate the MD simulation cell.

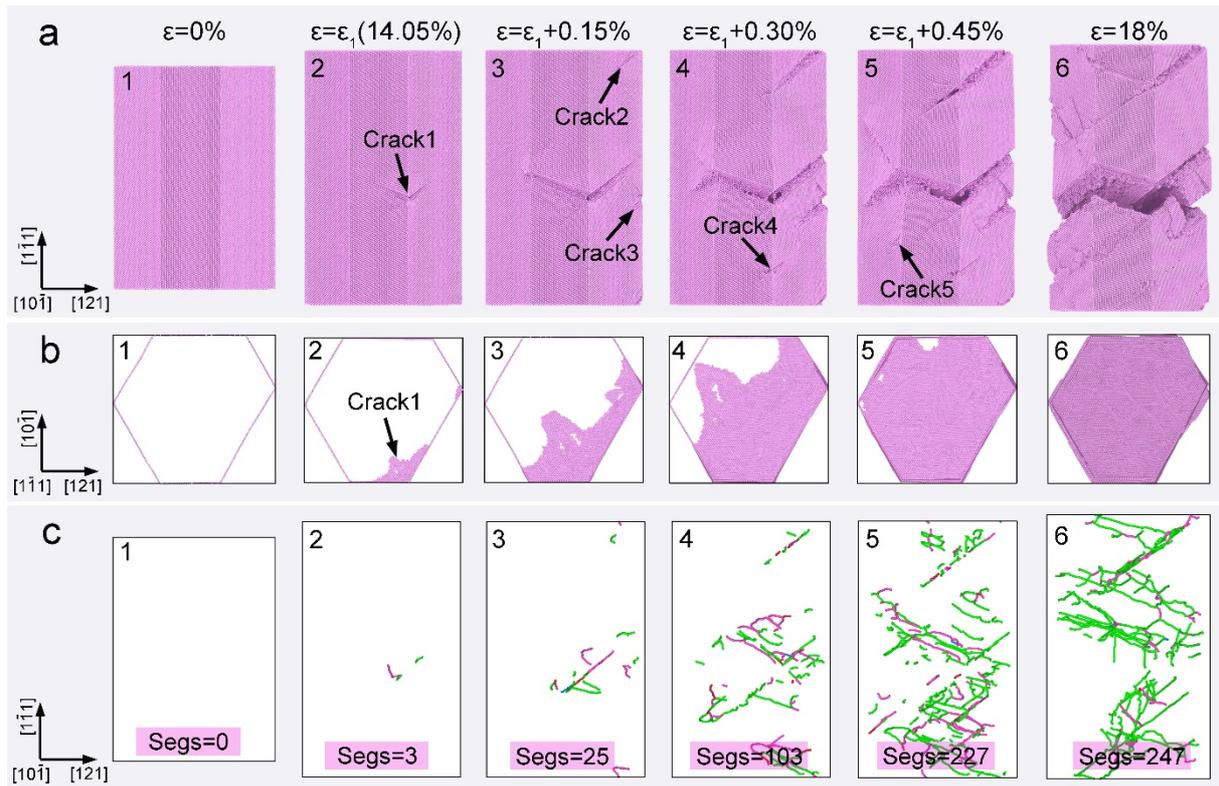

**Fig. 9.** Crack nucleation and propagation in a single-crystal W NW in tension along [1-11] direction. (a and b) Atomic model of the single-crystal W NW with a diameter of 21.8 nm viewed along the [10-1] direction (a) and the [1-11] direction (b) at strains of 0% (1), $\varepsilon_1$ = 14.5% (2), $\varepsilon_1$+0.15% (3), $\varepsilon_1$+0.30% (4), $\varepsilon_1$+0.45% (5), and 18% (6). $\varepsilon_1$ stands for the strain that the crack initiation. Atoms in the perfect BCC lattice are removed for clarity in (b). (**c**) Dislocation configurations in the single-crystal NW in (a) viewed along the [10-1] direction. Green, pink, blue and red lines represent 1/2<111>, <100>, <110> and other type dislocations. The number of dislocation segments (Segs) is 0 (1), 3 (2), 25 (3), 103 (4), 227 (5), and 247 (6). The black frames in (b) and (c) indicate the MD simulation cell.